\documentclass[ aps,%
12pt,%
floatfix,%
final,%
notitlepage,%
oneside,%
onecolumn,%
nobibnotes,%
nofootinbib,%
superscriptaddress,%
noshowpacs,%
]%
{revtex4}
\usepackage{amssymb}
\usepackage{caption}

\usepackage{pstricks,pst-node}
\psset{linewidth=.5pt,dash=4pt 4pt,unit=1cm,arrowsize=2pt 3}

\def\ArrowLine(#1,#2)(#3,#4){%
  \pcline(#1,#2)(#3,#4)%
  \lput{:U}{
    \pspicture(0,0)(0,0)
      \psline[arrows=->](2.3pt,0)(2.4pt,0)
    \endpspicture
  }
}

\begin{document}

\title{Possible nature of $Z^+(4430)$}
\author{S.S. Gershtein}
\affiliation{Institute for High Energy Physics, Protvino, Russia}
\author{A.K. Likhoded}
\affiliation{Institute for High Energy Physics, Protvino, Russia}
\author{G.P. Pronko}
\affiliation{Institute for High Energy Physics, Protvino, Russia}
\affiliation{Institute of Nuclear Physics, National Research Center "Demokritos", Athens, Greece}

\begin{abstract}
We discuss exotic states $X(3872)$ and $Z^+(4430)$, observed recently in experiment Belle. The QCD-string based explanation is suggested.
\end{abstract}
\maketitle

In addition to large number of $(c\bar c)$ mesons new "charmonium-like" exotic states appeared recently. The most exotic among them are $X(3872)$-meson \cite{R1} and resonant-like state $Z(4430)^+$ \cite{R2}. The discussion of $X(3872)$ properties can be found in large list of works \cite{R3}. Actually, there are two possibilities that are discussed in these works: 1) molecular nature of $X(3872)$, that is caused by strong interaction of $D$-mesons near threshold \cite{R4} and 2) 4-quark nature of this state (including diquark picture) \cite{R5}.

Leaving aside the first possibility we will try to give some arguments in favour of second variant. Let us consider the kinematics of weak $B$-meson decay $B\to K+X_{c\bar c}$, where $X_{c\bar c}$ is a final $c\bar c$-state and arbitrary amount of light quarks. In a simplest case, when $c\bar c$-pair is in colourless state, it forms charmonium levels $\eta_c$, $\psi$, $\psi'$, $\chi_{c0}$, $\chi_{c1}$ with close production rates. Comparable probabilities of production of these states in $B\to(c\bar c) K$ decay show, that their production proceeds at the scale typical for light hadrons $\sim 1$ fm. If these were not the case, one would observe a strong dependence of decay branching fractions on the properties of wave functions on small distances. In addition, the property of $b\to c\bar c s$ decay kinematics is that in interval $2.5\,\mathrm{GeV}<m_{c\bar c}<3.8\,\mathrm{GeV}$ the dependence on $c\bar c$ mass is weak.

The situation is completely different if $c\bar c$ pair is produced in colour-octet state (it should be noted, that this configuration is dominant). In this case, as it can be seen from Fig. 1, \begin{center}
\begin{pspicture}(0,0)(6.5,5.5)
\rput{30}(5.6,1){\psellipse[fillstyle=solid,fillcolor=lightgray](0,0)(1.6,1.2)}
\ArrowLine(5,0)(0,0)
\ArrowLine(0,2)(2.5,2)
\ArrowLine(6,1.5)(2.5,2)
\ArrowLine(2.5,2)(5,.5)
\ArrowLine(2.5,2)(6,5)
\qdisk(2.5,2){.05}
\pscurve(6,4.8)(4.5,3.2)(4.5,2.5)(6,1.7)
\psline[arrows=->](4.43,2.8)(4.43,2.7)
\rput(1.25,2.5){$b$}
\rput(6.3,5.2){$s$}
\rput(6.3,4.7){$\bar{q'}$}
\rput(6.3,1.9){$q'$}
\rput(6.3,1.4){$c$}
\rput(5.3,.5){$\bar c$}
\rput(1.25,0.3){$\bar q$}
\end{pspicture} \\[1em]
Fig.~1
\end{center}
after $s$-quark hadronization into $K$-meson the remaining $X_{c\bar c}$ system consists of four quarks. Further evolution of this system follows the well known scenario of production of two particles with open charm.  In the same time we can not exclude the possibility of production of compact 4-quark system with characteristic size $\sim 1$ fm, where the $c\bar c$-pair is in colour-octet state and has small invariant mass $\sim 2m_c$. In this paper we want to point out that this configuration of quarks arises naturally in standard approach, which describes ordinary (non-exotic) quark states $q\bar q$ and $qqq$ in string representation of long distance quark interaction. In this case we use to say, that quarks are bounded by strings of chromo-electric field. This kind of representation of ordinary hadrons is shown in Fig. 2.
\begin{center}
\begin{pspicture}(-.2,-1)(8.2,1.7)
\ArrowLine(0,0)(2,0)
\qdisk(0,0){.1}
\pscircle[fillstyle=solid,fillcolor=white](2,0){.1}
\pscircle(2,0){.1}
\rput(0,-.4) {$q$}
\rput(2,-.4) {$\bar q$}
\ArrowLine(7,1)(7,0)
\ArrowLine(6.2,-.5)(7,0)
\ArrowLine(7.8,-.5)(7,0)
\pscircle[fillstyle=solid,fillcolor=black](7,1){.1}
\pscircle(7,1){.1}
\pscircle[fillstyle=solid,fillcolor=black](6.2,-.5){.1}
\pscircle(6.2,-.5){.1}
\pscircle[fillstyle=solid,fillcolor=black](7.8,-.5){.1}
\pscircle(7.8,-.5){.1}
\rput(7,1.4){$q$}
\rput(5.9,-.8){$q$}
\rput(8.1,-.8){$q$}
\end{pspicture} \\[1em]
Fig. 2
\end{center}
The arrows on Fig. 1 show the direction of nonabelian chromoelectric flux. The presentation 3 comes out from quark $q$ and enters antiquark $\bar q$. Equivalent description corresponds  the representation $\bar 3$ enters $q$ and comes out from $\bar q$. In nonrelativistic limit for heavy quarks this description gives rise to the linear part of confining potential.

Is it is possible to generalize this description for more complicated 4-quark states? The answer is positive. In accordance with the representation theory of $SU_{col}(3)$ the representation 3 --- $\psi^\alpha$ is equivalent to the representation $\bar 3\otimes\bar 3$ --- $\psi_{[\alpha\beta]}$, antisymmetric with respect to indices $\alpha$ and $\beta$. Graphic representation (see Fig. 2)
is changed in the following way: the flux 3 either comes out from quark or two fluxes $\bar 3$ enter into the quark, correspondingly for antiquark we have to change $3\leftrightarrows\bar 3$ (see Fig. 3).
\begin{center}
\begin{pspicture}(-.5,-2)(8,2)
\ArrowLine(0,0)(2,0)
\pscircle[fillstyle=solid,fillcolor=black](0,0){.1}
\pscircle(0,0){.1}
\ArrowLine(2,-2)(0,-2)
\pscircle[fillstyle=solid,fillcolor=white](0,-2){.1}
\pscircle(0,-2){.1}
\ArrowLine(4,0)(6,0)
\ArrowLine(8,0)(6,0)
\pscircle[fillstyle=solid,fillcolor=black](6,0){.1}
\pscircle(6,0){.1}
\ArrowLine(6,-2)(4,-2)
\ArrowLine(6,-2)(8,-2)
\pscircle[fillstyle=solid,fillcolor=white](6,-2){.1}
\pscircle(6,-2){.1}
\rput(0,.5){$q$}
\rput(0,-1.5){$\bar q$}
\rput(6,.5){$q$}
\rput(6,-1.5){$\bar q$}
\end{pspicture} \\[1em]
Fig. 3
\end{center}
We want to emphasize, that this representation is not a novelty, but an accessor of ordinary QCD. This picture immediately implies the possibility of existence of compact $\sim 1$ fm four-quark configuration. Graphs corresponding to such states are presented in Fig. 4.
Note, that the state which correspond to Fig. 4a could not be reduced to simple picture of diquark interaction, while the state on the figure 4b may be interpreted this way. Also, as in our construction we use the elementary strings, which bear the flux of chromoelectic field 3 or $\bar 3$, the physical properties of such strings coincide with the ones in usual mesons. The other important issue is that any $q\bar q$ pair on the graph of Fig. 4 is in octet state.
\begin{center}
\begin{pspicture}(-.5,-1.5)(10.5,1.5)
\ArrowLine(2,1)(0,1)
\ArrowLine(2,1)(2,-1)
\ArrowLine(0,-1)(2,-1)
\ArrowLine(0,-1)(0,1)
\pscircle[fillstyle=solid,fillcolor=black](0,1){.1}
\pscircle(0,1){.1}
\pscircle[fillstyle=solid,fillcolor=white](2,1){.1}
\pscircle(2,1){.1}
\pscircle[fillstyle=solid,fillcolor=black](2,-1){.1}
\pscircle(2,-1){.1}
\pscircle[fillstyle=solid,fillcolor=white](0,-1){.1}
\pscircle(0,-1){.1}
\rput(-.3,1.3){$c$}
\rput(2.3,1.3){$\bar c$}
\rput(2.3,-1.3){$q$}
\rput(-.3,-1.3){$\bar q$}
\ArrowLine(4,0)(6,0)
\ArrowLine(8,0)(6,0)
\ArrowLine(8,0)(10,0)
\pscircle[fillstyle=solid,fillcolor=black](4,0){.1}
\pscircle(4,0){.1}
\pscircle[fillstyle=solid,fillcolor=black](6,0){.1}
\pscircle(6,0){.1}
\pscircle[fillstyle=solid,fillcolor=white](8,0){.1}
\pscircle(8,0){.1}
\pscircle[fillstyle=solid,fillcolor=white](10,0){.1}
\pscircle(10,0){.1}
\rput(4,-.4){$q$}
\rput(6,-.4){$c$}
\rput(8,-.4){$\bar c$}
\rput(10,-.4){$\bar q$}
\rput(1,-2){$a$}
\rput(7,-2){$b$}
\end{pspicture} \\[1em]
Fig. 4
\end{center}
Recall, that in the decay of $B$-meson we are discussing, the octet states are enhanced for $c\bar c$ pair. If we assume, that the $\mathrm{Br}(X(3872)\to\psi\pi\pi)$ and $\mathrm{Br}(Z^+(4430)\to\psi'\pi)$ are of order of several percents, we shall conclude, that the probability of production of these states is one of the order of probability of ordinary quarkonium production.

Note that the different structure of 4-quark states presented in Fig. 4 leads to different picture of their decay. If, for example, for decay of usual mesons (e.g. $\rho\to\pi\pi$) it is enough to consider only the breaking of string and production of $q\bar q$ pair, it is not possible for configuration shown in Fig. 4a. In order for this configuration to decay into $D\bar D$ or $\phi'\psi$ it is imperative to consider string interaction, which leads to their reconnection, as it is shown on figure 5. As far as the configuration on figure 4b is concerned, it may decay as usual hadrons into baryon-antibaryon pair. Unfortunately, because of low masses of the states $X(3872)$ and $Z^+(4430)$, this channel is not possible, but for higher excitations of these states it will dominate.
\begin{center}
\begin{pspicture}(-.5,-2)(14.5,2)
\ArrowLine(0,-1)(0,1)
\ArrowLine(2,1)(2,-1)
\pscurve(0,1)(.5,.5)(1.5,.5)(2,1)
\psline[arrows=->](1.05,.4)(.95,.4)
\pscurve(0,-1)(.5,-.5)(1.5,-.5)(2,-1)
\psline[arrows=->](.95,-.4)(1.05,-.4)
\pscircle[fillstyle=solid,fillcolor=black](0,1){.1}
\pscircle(0,1){.1}
\pscircle[fillstyle=solid,fillcolor=white](2,1){.1}
\pscircle(2,1){.1}
\pscircle[fillstyle=solid,fillcolor=black](2,-1){.1}
\pscircle(2,-1){.1}
\pscircle[fillstyle=solid,fillcolor=white](0,-1){.1}
\pscircle(0,-1){.1}
\rput(-.3,1.3){$c$}
\rput(2.3,1.3){$\bar c$}
\rput(2.3,-1.3){$q$}
\rput(-.3,-1.3){$\bar q$}
\ArrowLine(4,-1)(4,1)
\ArrowLine(6,1)(6,-1)
\pscurve(4,1)(4.5,.13)(5.5,.13)(6,1)
\pscurve(4,-1)(4.5,-.13)(5.5,-.13)(6,-1)
\psline[arrows=->](4.18,.545)(4.17,.565)
\psline[arrows=->](4.17,-.565)(4.18,-.545)
\psline[arrows=->](5.83,.565)(5.82,.545)
\psline[arrows=->](5.82,-.545)(5.83,-.565)
\pscircle[fillstyle=solid,fillcolor=black](4,1){.1}
\pscircle(4,1){.1}
\pscircle[fillstyle=solid,fillcolor=white](6,1){.1}
\pscircle(6,1){.1}
\pscircle[fillstyle=solid,fillcolor=black](6,-1){.1}
\pscircle(6,-1){.1}
\pscircle[fillstyle=solid,fillcolor=white](4,-1){.1}
\pscircle(4,-1){.1}
\rput(3.7,1.3){$c$}
\rput(6.3,1.3){$\bar c$}
\rput(6.3,-1.3){$q$}
\rput(3.7,-1.3){$\bar q$}
\ArrowLine(8,-1)(8,1)
\ArrowLine(10,1)(10,-1)
\pscurve(8,-1)(8.5,-.5)(8.5,.5)(8,1)
\pscurve(10,-1)(9.5,-.5)(9.5,.5)(10,1)
\psline[arrows=->](8.59,-.01)(8.59,.01)
\psline[arrows=->](9.41,.01)(9.41,-.01)
\pscircle[fillstyle=solid,fillcolor=black](8,1){.1}
\pscircle(8,1){.1}
\pscircle[fillstyle=solid,fillcolor=white](10,1){.1}
\pscircle(10,1){.1}
\pscircle[fillstyle=solid,fillcolor=black](10,-1){.1}
\pscircle(10,-1){.1}
\pscircle[fillstyle=solid,fillcolor=white](8,-1){.1}
\pscircle(8,-1){.1}
\rput(7.7,1.3){$c$}
\rput(10.3,1.3){$\bar c$}
\rput(10.3,-1.3){$q$}
\rput(7.7,-1.3){$\bar q$}
\ArrowLine(12,1)(12,-1)
\ArrowLine(14,-1)(14,1)
\pscircle[fillstyle=solid,fillcolor=black](12,1){.1}
\pscircle(12,1){.1}
\pscircle[fillstyle=solid,fillcolor=white](14,1){.1}
\pscircle(14,1){.1}
\pscircle[fillstyle=solid,fillcolor=black](14,-1){.1}
\pscircle(14,-1){.1}
\pscircle[fillstyle=solid,fillcolor=white](12,-1){.1}
\pscircle(12,-1){.1}
\rput(11.7,1.3){$c$}
\rput(14.3,1.3){$\bar c$}
\rput(14.3,-1.3){$q$}
\rput(11.7,-1.3){$\bar q$}
\psline[arrows=->](2.5,0)(3.5,0)
\psline[arrows=->](6.5,0)(7.5,0)
\psline[arrows=->](10.5,0)(11.5,0)
\end{pspicture} \\[1em]
Fig. 5
\end{center}
Let us summarize our qualitative discussion. The states $X(3872)$ and $Z^+(4430)$ have bright indication of exotic states. The dominant mode of $X(3872)$ decay is into $D^0\bar D^{0*}$. The state $Z^+(4430)$ has sufficiently big width also due to the decay into $D^{(*)}\bar D^{(*)}$. Allowed quantum numbers of $\psi'\pi^+$ system is $0^{-}$, $1^-$, $0^+$ and $2^-$ though the last state is most probable suppressed in weak decay of $B$-meson. The resonance $Z^+(4430)$ is on the vicinity of threshold $D_{1^-}^{*}(2010) D_{1^+}^{*}(2432)$. This fact led the authors of \cite{R6} to interpretation of $Z^+(4430)$ as threshold singularity. Apparently this is possible if the quantum numbers of $\psi'\pi^+$ are $0^-$ or $1^-$. If in this case $Z^+(4430)$ is a resonant state, its total width should be saturated by the channel $Z^+(4430) \to D^*_{1^-}\bar D^*_{1-}$, in the same time the problem of the channel $Z^+\to\psi'\pi$ is still unsolved.

The very existance of tetraquark states in QCD opens many possibilities of observation of exotic states in the same decay of $B$-meson. For example, if we change the light pair of quarks on figure 1 to $s\bar s$, we obtain the process of production of strange tetraquarks $Z_s$ and $X_s$ with the mass bigger than $Z$ and $X$ masses by $\Delta M=m_s-m_q \approx 0.15$ GeV in the decay channel
\begin{center}
\begin{pspicture}(0,-1)(3.5,.5)
\rput(0,0){$B$}
\psline[arrows=->](.3,0)(.7,0)
\rput(1,0){$\phi$}
\rput(1.5,0){$Z_s$}
\psline(1.5,-.3)(1.5,-.7)
\psline[arrows=->](1.5,-.7)(2.3,-.7)
\rput(2.9,-.7){$\psi' K$}
\end{pspicture}
\end{center}
The dominant mode for $Z_s$ as in decays of $Z^+(4430)$ should be $D^* D^*_s$.


Further, if we change on Fig.~1 $\bar q$ onto the pair $q_1 q_2$, we obtain the diagram of decay of b-baryon into the $K$-meson and pentaquark with hidden charm $\Pi_{c\bar c}$. Corresponding graph for pentaquark is shown on Fig. 6.

\begin{center}
\begin{pspicture}(0,-1)(8,.5)
\ArrowLine(0,0)(2,0)
\ArrowLine(4,0)(2,0)
\ArrowLine(4,0)(6,0)
\ArrowLine(8,0)(6,0)
\pscircle[fillstyle=solid,fillcolor=black](0,0){.1}
\pscircle(0,0){.1}
\pscircle[fillstyle=solid,fillcolor=black](2,0){.1}
\pscircle(2,0){.1}
\pscircle[fillstyle=solid,fillcolor=white](4,0){.1}
\pscircle(4,0){.1}
\pscircle[fillstyle=solid,fillcolor=black](6,0){.1}
\pscircle(6,0){.1}
\pscircle[fillstyle=solid,fillcolor=black](8,0){.1}
\pscircle(8,0){.1}
\rput(0,-.4){$q$}
\rput(2,-.4){$q$}
\rput(4,-.4){$\bar c$}
\rput(6,-.4){$c$}
\rput(8,-.4){$q$}
\end{pspicture} \\[1em]
Fig. 6
\end{center}

For decay chain $\Lambda_b\to\Pi_{c\bar c}K$ we expect, analogously to $Z^+(4430)$ decay, the decay $\Pi_{c\bar c}\to\psi'N$.

We thank A.V. Razumov and A.V. Luchinsky. This work was supported by Russian foundation for basic research, grants \# 07-02-00417, \# 07-01-00234.

\end{document}